\documentclass[aps,twocolumn]{revtex4}

\usepackage{xcolor}
\usepackage{graphicx}
\usepackage{amsmath,amssymb,amsfonts}
\newcommand{\be}{\begin{equation}}
\newcommand{\ee}{\end{equation}}
\newcommand{\ba}{\begin{eqnarray}}
\newcommand{\ea}{\end{eqnarray}}
\newcommand{\ban}{\begin{eqnarray*}}
\newcommand{\ean}{\end{eqnarray*}}

\begin{document}

\author{Berndt M\"uller}
\affiliation{Department of Physics, Duke University, Durham, NC 27708-0305, USA}
\author{Andreas Sch\"afer}
\affiliation{Institut f\"ur Theoretische Physik, Universit\"at Regensburg, D-Regensburg, Germany}
\title{Why does the thermal model for hadron production in heavy ion collisions work?}

\begin{abstract}
The yields for hadrons and even light nuclei measured at midrapidity in relativistic heavy ion collisions are found to be dictated exclusively by their thermal Boltzmann factor for a common temperature of approximately 155 MeV. The reason for the validity of the ``thermal model'' description is widely discussed. Here, we offer a new type of argument in its favor. 
\end{abstract}

\maketitle

\section{Introduction}

The standard paradigm for high energy heavy ion collisions is that colliding nuclei form a fireball that cools by expanding and finally hadronizes into individual uncorrelated hadrons \cite{Westfall:1976fu}. This paradigm is in fundamental conflict with the fact that QCD is time reversal invariant and thus QCD interactions cannot create entropy for an isolated system like that formed by the colliding nuclei in the beam-line vacuum. Time reversal invariance, combined with unitary time evolution of isolated quantum states thus implies a significant entanglement of the wave functions of the produced hadrons. 

This argument provides strong motivation to look for anomalies in heavy ion collisions that could indicate that the assumption of a incoherent, thermal hadron gas that underlies the ``thermal model'' \cite{Andronic:2016nof} fails. Instead, the ALICE experiment has shown that even light, weakly bound nuclei like  the hypertriton $^3_{\Lambda}H$ are produced exactly as predicted by a thermal Boltzmann distribution with the same temperature as all tightly bound baryons, such as the proton. This observation appears incompatible with the intuitive picture of a thermal hadron gas in which any weakly bound nuclear state should immediately break-up due to collisions with other particles, or get produced only at a much lower freeze-out temperature.  Compounding the puzzle, the common temperature for all produced hadrons and nuclei is found to be the QCD phase transition temperature (see Fig.4 in \cite{Andronic:2016nof}), not a lower freeze-out temperature. These anomalies are so striking that the consideration of unconventional descriptions of heavy ion collisions is well motivated. 

The idea that entanglement and entanglement entropy should play a crucial role in the dynamics of heavy ion collisions is not new. An example can be found in \cite{Muller:2011ra}, where the present authors argued that the manifestation by decoherence of the entanglement entropy encoded in the wave functions of the colliding nuclei can explain a sizable fraction of the apparent entropy of the final state. Similarly, Ho and Hsu \cite{Ho:2015rga} argued that quantum entanglement of an observed subsystem with the entire system can mimic thermalization, and that the entaglement entropy between subsystems increases rapidly with time. Two very recent publications \cite{Kharzeev:2017qzs,Berges:2017zws} have linked the concept of entanglement entropy to the phenomenology of heavy ion collisions in other specific ways. The present article pursues a similar line of arguments. 

\section{Unitary evolution and entanglement}

As long as information about the state of a time reversal invariant quantum system is not lost by any kind of measurement and the associated (partial) collapse of the complete many-body wave function, it must be possible ``to run the film backwards''. Thus we can argue that a highly complex initial state $\Phi_\mathrm{i}$ characterized by a parton density matrix with two blocks describing the two nuclei approaching one another evolves by a unitary transformation into another highly entangled final state $\Phi_\mathrm{f}$ characterized by a full density matrix in the parton basis. The final many-parton quantum state is then projected onto hadron states by experimental measurements. 

An empirical property of heavy-ion collisions is that at intermediate times a state $\Phi_\mathrm{th}$ forms at midrapidity for which few-particle observables are to very good approximation described by their values in a thermal ensemble. This intermediate state is called the ``fireball''. For the same reason as given above, namely conservation of information, this state cannot really be a truly thermal state, but for few-parton observables it can look thermal to a very good approximation, as will be discussed below.

The time evolution of the isolated system is described by the unitary QCD time evolution operator $\hat U$
\begin{equation}
|\Phi_\mathrm{f}\rangle  = \hat U(t_\mathrm{f},t_\mathrm{th}) ~\hat U(t_\mathrm{th},t_\mathrm{i}) |\Phi_\mathrm{i}\rangle = \hat U(t_\mathrm{f},t_\mathrm{th})~|\Phi_\mathrm{th}\rangle ,
\end{equation}
where we have split the full time evolution into the evolution from the incident nuclei to the  quasi-thermal intermediate fireball state, and the evolution from the fireball state to the time at which the measurement in the detector takes place.  Any measurement of a hadron or nucleus $N$ at time $t_\mathrm{f}$ corresponds to the projection of $|\Phi_\mathrm{f}\rangle$ with some projection operator $\hat{\cal P}_N$. We are free to define this projection operator for any other time $t$ as
\begin{eqnarray}
\hat {\cal P}_N (t) &=& \sum_{X_N} ~|N\rangle \langle N| ~ \otimes ~|X_N(t)\rangle \langle X_N(t)|
\nonumber \\
&=&|N\rangle \langle N| ~\otimes~ \sum_{X_N}~|X_N(t)\rangle \langle X_N(t)|
\end{eqnarray}
with the QCD eigenstate $|N\rangle$. This decomposition is possible as long as the latter can exist, i.e.\ for any time after the break-up time $t_\mathrm{bu}$. $|X_N(t)\rangle$ is here any state of the colliding system without $N$, i.e. any state with 4-momentum $p^{\mu}({\rm nucleus~1})+p^{\mu}({\rm nucleus~2})-p^{\mu}(N)$ and with all corresponding conserved charges given by the sum for both nuclei minus the value for $N$.  We emphasize that because $\hat {\cal P}_N (t)$ is only constrained by its late time limit, it is not unique. Our choice, which sets the non-interacting QCD eigenstate $|N\rangle$ apart, is motivated by the peculiar empirical properties of the QCD fireball. Without these non-trivial properties our choice would not be helpful.

Because $|N\rangle$ is a QCD eigenstate, the time evolution of ${\cal P}_N $ is given by
\begin{eqnarray}
 &&\hat U(t,t')^{\dagger}\hat {\cal P}_N (t)  \hat U(t,t')~=~ \sum_{X_N} ~e^{{-\rm i}E_N(t-t')}|N\rangle \langle N| 
\nonumber \\
&\otimes&  
\sum_{X_N}~\hat U(t,t')^{\dagger}|X_N(t)\rangle \langle X_N(t)| \hat U(t,t')
\end{eqnarray}
While the explicit action of $\hat {\cal P}_N (t)$ on the colliding heavy ion system is, in general, intractable, there is one moment in time when this is not true: Quasi-thermalization and quasi-hydrodynamization \cite{Romatschke:2017vte,Florkowski:2017olj,Romatschke:2017acs} imply that single particle observables in the fireball and at the time of its break-up $t_\mathrm{bu}$ behave as if the system would be thermal. In our notation this means that the probability  
\begin{equation}
\mathrm{Tr} \Bigl\{ \hat{\cal P}_N(t_\mathrm{bu})  |\Phi_\mathrm{bu}\rangle \langle \Phi_\mathrm{bu}|\hat{\cal P}_N(t_\mathrm{bu}) \Bigr\}
\end{equation}
can be calculated using fireball probability distributions for quarks and gluons neglecting entanglement effects. While the empirical fact of particle yields following the thermal description seems to be indisputable, the theoretical understanding of quasi-thermalization and quasi-hydrodynamization is still much debated.   

The probability to form a state $N$ at time $t_\mathrm{f}$ is given by the trace over $X_N$ of the projected density matrix
\begin{eqnarray} 
&& \mathrm{Tr}_{X_N} \Bigl\{ \hat{\cal P}_N(t_\mathrm{f}) |\Phi_\mathrm{f}\rangle \langle \Phi_\mathrm{f}| \hat{\cal P}_N(t_\mathrm{f}) ~ \Bigr\}
\nonumber \\
&=& \mathrm{Tr}_{X_N} \Bigl\{  \hat U(t_\mathrm{f},t_\mathrm{bu})~\hat{\cal P}_N (t_\mathrm{bu}) \hat U(t_\mathrm{bu},t_\mathrm{f})^{\dagger} \hat U(t_\mathrm{f},t_\mathrm{bu}) |\Phi_\mathrm{bu}\rangle 
\nonumber \\
&\times& \langle \Phi_\mathrm{bu}| 
\hat U(t_\mathrm{f},t_\mathrm{bu})^{\dagger} \hat U(t_\mathrm{bu},t_\mathrm{f}) \hat{\cal P}_N(t_\mathrm{bu}) \hat U(t_\mathrm{bu},t_\mathrm{f})^{\dagger}   \Bigr\}
\nonumber \\
&=& \mathrm{Tr} \Bigl\{ \hat{\cal P}_N(t_\mathrm{bu})  |\Phi_\mathrm{bu}\rangle \langle \Phi_\mathrm{bu}|\hat{\cal P}_N(t_\mathrm{bu}) \Bigr\}~~~.
\label{eq:Tr}
\end{eqnarray}

\begin{figure}[htb]      
\centering
\includegraphics[width=0.95\linewidth]{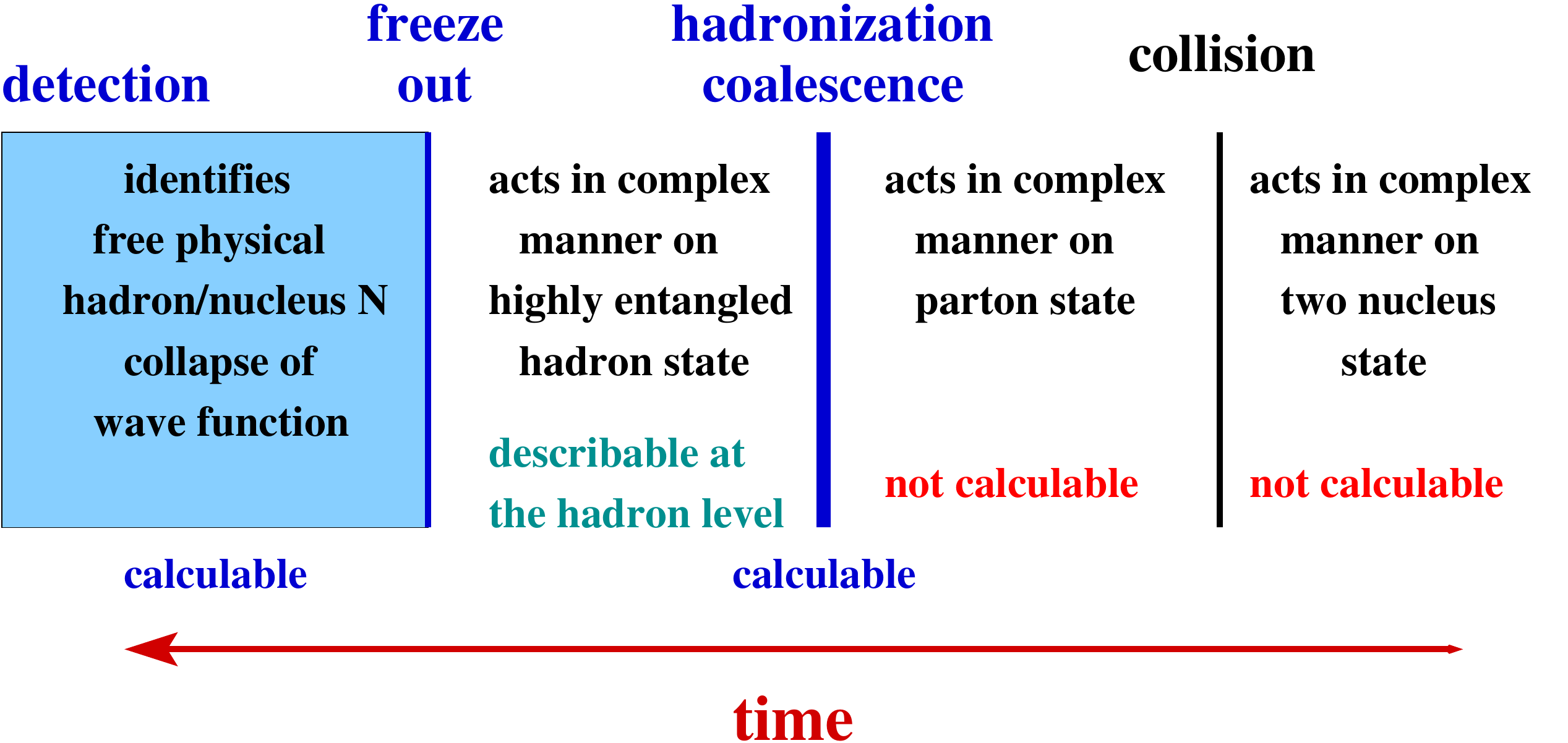}
\caption{The action of $\hat {\cal P}_N (t)$: Schematic representation of the different stages of the collision. The projection on an entangled hadronic eigenstate is only calculable in a simple manner at two moments: at break-up of the quasi-thermal parton fireball and when the state is measured in the detector at late times.}
\label{fig1}
\end{figure}

The essence of our argument is that, because QCD is time reversal invariant, the probability for forming any asymptotic hadron state 
can, in principle, be evaluated at any time. However, in practice, this is impossible except for late times, i.e. in the asymptotic or near asymptotic hadronic phase and at the break-up time $t_\mathrm{bu}$ of the quark-gluon plasma. We will pursue the second alternative in the following.  The sequence of the different stages of the collision system and the action of the projector $\hat{\cal P}_N$ are schematically illustrated in Fig.~\ref{fig1}.

\section{Digression: The information paradox}

To motivate the use of thermal distributions at this intermediate stage, although quantum entanglement still exists, we envision this thermal state to be very similar to the (quasi-thermal) Hawking radiation emitted by a black hole. One of the deep questions of black hole physics is whether black hole formation and decay are described by a unitary, {\em i.e.}, information preserving operator and, if so, how all the information contained in the initial state is stored in the final one. A widely accepted interpretation is that the process is indeed adiabatic, {\em i.e.}, entropy and thus information conserving, and that the information is stored in the final state in entanglement phases among all photons emitted in the form of Hawking radiation. The information would then not be lost, in principle, but could not be retrieved in practice. As a consequence, different arguments can lead to seemingly contradictory conclusions, which is why this discussion is referred to as ``black hole information paradox''. The time evolution of the complete quantum system would be unitary, but all few-photon observables in the final state would appear thermal to a very high degree. We posit that the same concept applies to heavy ion collisions.   

One might even be tempted to link this analogy to the AdS/CFT duality, in which thermalization in the conformal boundary field theory (CFT) corresponds to black hole formation in the bulk. However, when doing so one encounters the limits of this duality. A conformal theory cannot be confining, thus, the fireball continues to expand and cool without transitioning to a hadron gas. On the string theory side of the duality the formed AdS black hole/black brane is stable, in contrast to a Minkowski black hole. Standard AdS/CFT analysis (see {\em e.g.} \cite{Chesler:2008hg,vanderSchee:2013pia,Buchel:2016cbj,Balasubramanian:2010ce}) is thus the method of choice to study {\em qualitative} features of the early dynamics of heavy ion collisions, but is not applicable to their late time development. 

As reviewed in \cite{Rebhan:2014rxa} the Sakai-Sugimoto model variant of a gravity dual theory, which includes confinement and can thus describe hadronization and reproduce empirical hadron masses and decay constants, could allow for a more complete and quantitative description of heavy ion collisions. In a series of publications, Mandal, Morita, and Isono \cite{Mandal:2011ws,Mandal:2011uq,Isono:2015uda} have worked out a consistent formalism to treat QCD thermodynamics within the Sakai-Sugimoto model. However, the technical difficulties of performing numerical calculations for realistic heavy ion collisions, like those of \cite{Chesler:2015wra} for the standard AdS/CFT theory, are much more severe for the Sakai-Sugimoto model, and such calculations have not yet been performed. 

Another analogy which may help elucidate the issue is to compare the time evolution of a heavy ion collision to calculations by a quantum computer. A standard starting point for many discussions in quantum computing is the observation that the action of an ideal quantum computer can be described as the unitary time evolution of an isolated physical system under a time reversal invariant Hamiltonian. For example, because ``information is physical'' \cite{Landauer} the laws of thermodynamics also apply to quantum computing. Here we invert this argument by stating that the time evolution caused by QCD interactions transforms entangled quantum states without any information loss just like a quantum computer. Information loss occurs only when data are read out or when the system couples in some other way to an external heat bath as happens in QCD during the interaction of the emitted particles with a detector. Therefore, the assumption of a decohered state at intermediate stages is potentially misleading. Before any measurement takes place, all hadron wave functions are highly entangled and any precise calculation would need to take the resulting interference and correlation effects into account. We argue, however, that because of the very special properties of the quark-gluon plasma such a simplifying assumption is justified at the break-up time of the fireball.

\section{The role of unitarity}

Let us now continue our argument. We have tried to motivate why we can assume a thermal reservoir of parton states $p_\mathrm{i}(k_{\mu})$ for calculation of break-up probabilities. Let us further define $Q_N$ as the ensemble of all combinations of partons that can coalesce into the hadron or nucleus state $|N\rangle$.  (Here $N$ is assumed to be the only nuclear bound state for a given quantum number and invariant mass.)  The only other assumption we make is that to good approximation we can neglect the entanglement among these parton states as they form only a small subsystem of the fireball state $\Phi_\mathrm{th}$. Then, any state $\Psi_{Nj}\in Q_N$ with $M_{Nj}$ partons can be described as a simple product state of the parton wave functions except for the antisymmetrization of fermions, which we indicate by the fermion antisymmetrization operator $\hat F$.
\begin{eqnarray}
|\Psi_{Nj}\rangle &=& \hat F ~ |p_{NJ1}(k_{\mu}(1))\rangle ~ |p_{NJ2}(k_{\mu}(2))\rangle 
\nonumber \\
& & \ldots~ |p_{NJM}(k_{\mu}(M_{Nj}))\rangle 
\label{eq:PsiNj}
\end{eqnarray}
where for each state $[k_{\mu}(1)+k_{\mu}(2) ...k_{\mu}(M_{Nj})]^2=m_N^2$ is the invariant mass squared of the state $N$ and the flavor, charge and spin quantum numbers of all partons add up to those of $N$.

Next we expand the state $N$ in this parton basis 
\begin{eqnarray}
|N> &=& \sum_j ~ \sum_{s_1} \sum_{s_2}... \sum_{s_{M_{Nj}}}~\int d^3k(1) ... d^3k(M_{Nj})
\nonumber \\ 
&\times& |\tilde{\Psi}_{Nj}(k_\mu(1), k_{\mu}(2) ...k_{\mu}(M_{Nj})s_1, s_2 ...s_{M_{Nj}}))\rangle 
\nonumber \\
&\times& c_{Nj}(k_\mu(1), k_{\mu}(2) ...k_{\mu}(M_{Nj}),
s_1, s_2 ...s_{M_{Nj}})
\label{eq:cNj-1}
\end{eqnarray}
with the completeness relation
\begin{eqnarray}
1&=& \sum_{s_1} \sum_{s_2}... \sum_{s_{M_{Nj}}}~
\int d^3k(1) ... d^3k(M_{Nj})
\\
&\times&|c_{Nj}(k_\mu(1), k_{\mu}(2) ...k_{\mu}(M_{Nj}),s_1, s_2 ...s_{M_{Nj}})|^2
\nonumber
\label{eq:cNj-2}
\end{eqnarray}
where each state $\tilde{\Psi}_{Nj}$ is normalized to unity, while in the thermal state it is normalized according to the Boltzmann distribution, {\em i.e.}, 
\begin{eqnarray}
\langle \Psi_{Nj} |\Psi_{Nj} \rangle &=&  
e^{-(E_1+E_2+...+E_{M_{Nj}})/kT}~\langle \tilde\Psi_{Nj} |\tilde\Psi_{Nj} \rangle 
\nonumber \\
&=&  e^{-E_N/kT}
\end{eqnarray}
where we have approximated the Fermi and Bose distributions by Boltzmann factors. The observation that in a thermal ensemble the Boltzmann factors always combine to give the same factor $e^{-E_N/kT}$, independent of the number of coalescing partons, was already crucial for the arguments made in \cite{Muller:2005pv}.

Using orthogonality with respect to the $k(1), \ldots , k(M_{Nj})$ and $s_1, s_2, \ldots , s_{M_{Nj}}$ as well as (\ref{eq:PsiNj}, \ref{eq:cNj-1}) and (\ref{eq:cNj-2}), we obtain
\begin{equation}
|\langle N|\Psi_\mathrm{th}\rangle|^2 =  e^{-E_N/kT} ,
\end{equation}
which is just the result of the thermal model. 

Our reasoning does not exclude the presence of entanglement and non-trivial correlations within $|N\rangle$, {\em e.g.}, between two produced protons and neutrons in a ~$^4$He nucleus. As long as this state is an eigenstate of the QCD Hamiltonian, time evolution, which we denoted by $\hat U(t_\mathrm{f},t_\mathrm{th})$, includes the effect of the nucleon-nucleon interaction to all orders. The projection operator is that of the complete state, not the product of four single-nucleon projection operators. For that reason, time evolution for, {\em e.g.}, a~$^4$He nucleus includes all binding effects such that the energy that enters the Boltzmann factor is that for the physical mass including binding effects. 

We note parenthetically that hadrons containing heavy quark, such as heavy quarkonium states, must be treated differently as they can be produced at temperatures above the QCD hadronization temperature for light quarks by coalescence of heavy quarks that were pair produced in hard interactions among incident partons.  

\section{Correlations}

The reader may ask why two-particle correlations may reflect rescattering among produced particles at a later stage, when single particle yields do not. In order to pursue this question, we take entanglement into account. The usual discussion of Hanbury Brown-Twiss intensity interferometry, see, e.g.,  Eq.(2) to (5) from \cite{Baym:1997ce}, is then modified as follows.
The amplitude at detector 1 is:
\begin{equation}
A_1 =\frac{1}{L} \left( \alpha |A_1(t_1)\rangle e^{{\rm i} kr_{1a}+{\rm i}\phi_a}  
+ \beta |B_1(t_1)\rangle e^{{\rm i} kr_{1b}+{\rm i}\phi_b}\right)  
\end{equation}
Where $|A(t_1)\rangle$ and $|B(t_1)\rangle$ describe the rest of the entangled system wave function after the projection caused by measuring a particle, e.g. a pion, at time $t_1$. In general 
$|A(t_1)\rangle$ and $|B(t_1)\rangle$ differ. The corresponding probability is  
\begin{eqnarray}
I_1 &=& \frac{1}{L}~\Bigl( |\alpha|^2+ |\beta|^2 
\nonumber\\
&+& \alpha^*\beta\langle A_1(t_1)|B_1(t_1)\rangle 
e^{{\rm i}[ k(r_{1b}-r_{1a})+\phi_b-\phi_a]}
\nonumber \\
&+& \alpha\beta^*\langle B_1(t_1)|A_1(t_1)\rangle 
e^{{\rm i}[ k(r_{1a}-r_{1b})+\phi_a-\phi_b]}\Bigr)~~~.
\end{eqnarray}  
For the measurement by detector 2 one gets 
\begin{eqnarray}
I_2 &=& \frac{1}{L}~\Bigl( |\alpha|^2+ |\beta|^2 
\nonumber\\
&+& \alpha^*\beta\langle A_2(t_2)|B_2(t_2)\rangle 
e^{{\rm i}[ k (r_{2b}-r_{2a})}+\phi_b-\phi_a]
\nonumber \\
&+& \alpha\beta^*\langle B_2(t_2)|A_2(t_2)\rangle 
e^{{\rm i}[ k (r_{2a}-r_{2b})}+\phi_a-\phi_b]\Bigr)~~~.
\end{eqnarray}  
After event averaging the oscillatory parts vanish due to the presence of the phase $\phi_a-\phi_b$:
\begin{eqnarray}
\Big\langle I_1 \Big\rangle_\textrm{av} 
&=& \frac{1}{L^2}~\Bigl( |\alpha|^2+ |\beta|^2 \Bigr) ,
\end{eqnarray}
where $\langle \cdots \rangle_\textrm{av}$ indicates an event average. The product amplitude reads
\begin{eqnarray}
A_1 A_2 &=& \frac{1}{L^2} \Bigl( \alpha \alpha |A_1(t_1)\rangle e^{{\rm i} kr_{1a}+{\rm i}\phi_a}  
|A_2(t_2)\rangle e^{{\rm i} kr_{2a}+{\rm i}\phi_a}  
\nonumber \\
&+& \alpha \beta |A_1(t_1)\rangle e^{{\rm i} kr_{1a}+{\rm i}\phi_a}|B_2(t_2)\rangle e^{{\rm i} kr_{2b}+{\rm i}\phi_b}
\nonumber \\
&+& \alpha \beta |A_2(t_2)\rangle e^{{\rm i} kr_{2a}+{\rm i}\phi_a}|B_1(t_1)\rangle e^{{\rm i} kr_{1b}+{\rm i}\phi_b}
\nonumber \\
&+& \beta \beta |B_1(t_1)\rangle e^{{\rm i} kr_{1b}+{\rm i}\phi_b}  
|B_2(t_2)\rangle e^{{\rm i} kr_{2b}+{\rm i}\phi_b}\Bigr)  
\end{eqnarray}
We assume that the detection times within the two detectors are not equal within 
$1/k$ and that, therefore, all matrix elements of the type 
$\langle X(t_1)|Y(t_2)\rangle$ with $X,Y= A_1, A_2, B_1, B_2$ average to zero.
For the averaged product intensity we get then 
\begin{eqnarray}
\Big\langle I_1I_2 \Big\rangle_\textrm{av} &=& 
|\alpha|^2|\alpha|^2\langle A_1(t_1)|A_1(t_1)\rangle\langle A_2(t_2)|A_2(t_2)\rangle
\nonumber \\
&+&|\alpha|^2|\beta|^2\langle A_1(t_1)|A_1(t_1)\rangle\langle B_2(t_2)|B_2(t_2)\rangle
\nonumber \\
&+&|\alpha|^2|\beta|^2\langle B_1(t_1)|B_1(t_1)\rangle\langle A_2(t_2)|A_2(t_2)\rangle
\nonumber \\
&+&|\beta|^2|\beta|^2\langle B_1(t_1)|B_1(t_1)\rangle\langle B_2(t_2)|B_2(t_2)\rangle
\nonumber \\
&+&|\alpha|^2|\beta|^2\Big\langle\langle A_2(t_2)|B_2(t_2)\rangle\langle B_1(t_1)|A_1(t_1)\rangle
\nonumber\\
&\times& e^{{\rm i}k(r_{2b}-r_{2a}+r_{1a}-r_{1b})}\Big\rangle_\textrm{av}
\nonumber \\
&+&|\alpha|^2|\beta|^2\Big\langle\langle A_1(t_1)|B_1(t_1)\rangle\langle B_2(t_2)|A_2(t_2)\rangle
\nonumber\\
&\times& e^{{\rm i}k(r_{1b}-r_{1a}+r_{2a}-r_{2b})}\Big\rangle_\textrm{av}
\nonumber \\
&=&\Big\langle I_1\Big\rangle_\textrm{av} \Big\langle I_2 \Big\rangle_\textrm{av}
\nonumber \\
&+&|\alpha|^2|\beta|^2\Bigl(\Big\langle\langle A_2(t_2)|B_2(t_2)\rangle\langle B_1(t_1)|A_1(t_1)\rangle
\nonumber\\
&\times& e^{{\rm i}k(r_{2b}-r_{2a}+r_{1a}-r_{1b})}\Big\rangle_\textrm{av}
\nonumber \\
&+&\Big\langle\langle A_1(t_1)|B_1(t_1)\rangle\langle B_2(t_2)|A_2(t_2)\rangle
\nonumber\\
&\times& e^{{\rm i}k(r_{1b}-r_{1a}+r_{2a}-r_{2b})}\Big\rangle_\textrm{av} \Bigr) .
\end{eqnarray}
The matrix elements 
$\langle A_1(t_1)|B_1(t_1)\rangle\langle B_2(t_2)|A_2(t_2)\rangle$
and 
$\langle A_2(t_2)|B_2(t_2)\rangle\langle B_1(t_1)|A_1(t_1)\rangle$
average to zero if there is still noticable interactions between any of the final state 
hadrons. However, after freeze-out (note that the time of freeze out is the same 
for all hadrons due to the entanglement effects), when all dynamics stopped, we can assume
\begin{eqnarray}
\langle A_1(t_1)|B_1(t_1)\rangle &=& \langle A_2(t_2)|B_2(t_2)\rangle
\nonumber \\
\langle B_1(t_1)|A_1(t_1)\rangle &=& \langle B_2(t_2)|A_2(t_2)\rangle
\end{eqnarray}
and obtain 
\begin{eqnarray}
\Big\langle I_1I_2 \Big\rangle_\textrm{av} 
&=& \Big\langle I_1 \Big\rangle_\textrm{av} \Big\langle I_2 \Big\rangle_\textrm{av}
\nonumber \\
&+&2|\alpha|^2|\beta|^2 |\langle A_1(t_1)|B_1(t_1)\rangle|^2
\nonumber\\
&\times& \cos[k(r_{2b}-r_{2a}+r_{1a}-r_{1b})]
\end{eqnarray}
Thus it is explained quite naturally why HBT-correlations are dominated by freeze-out conditions. Note, however, that the factor  $|\langle A_1(t_1)|B_1(t_1)\rangle|^2$ could lead to corrections.

\section{Conclusions}

We conclude that there is good reason to expect the thermal model to reproduce the production probabilities for any type of hadron or light nucleus, and the common temperature to be the phase transition temperature and not a ``freeze-out'' temperature. These experimentally observed properties follow from the argument that, if a small part of the highly entangled fireball state is projected out, its probability distribution looks thermal. As such a behavior is characteristic for the {\em Eigenstate Thermalization Hypothesis} (ETH) \cite{DAlessio:2016rwt}, the success of the thermal model provides strong evidence that the ETH concept can be applied to the system formed by colliding heavy ions as already conjectured by Becattini and Fries \cite{Becattini:2009fv}.

We acknowledge valuable discussions with P.~Braun-Munzinger and S.~D.~H.~Hsu. This work was supported by DOE grant DE-FG02-05ER41367 and BMBF grant O5P15WRCCA. One of us (AS) thanks the Nuclear and Particle Physics Directorate of Brookhaven National Laboratory for its hospitality.

\end{document}